# The quantum description and stimulated emission radiation of coupled metamaterials

Shuming Wang[1], Hui Liu[1,*], Tao Li[1], Shining Zhu[1,*], Xiang Zhang[2]

*[1]National Laboratory of Solid State Microstructures and Department of Physics, Nanjing University, Nanjing, 210093, China*

*[2]5130 Etcheverry Hall, Nanoscale Science and Engineering Center, University of California, Berkeley, California 94720-1740, USA*

URL：http://dsl.nju.edu.cn/dslweb/images/plasmonics-MPP.htm

**Abstract**:

In recent years, there has been increasing interest in studying the quantum characteristics in plasmonic metamaterials. By using the Hamiltonian combined with second quantization, we have investigated the basic excitation of the coupled metamaterials and presented a quantum description of them. The interaction between the excitation of the coupled metamaterial and the exciton in the quantum dot material has been further considered with the aid of the interaction Hamiltonian. In such a damping system, a set of Langevin equations has been used to deal with the quantum motion of the system. And the stimulated emission has been found, which can be used to form a nano-laser. The quantum description of coupled metamaterial and the interaction system could work as the fundamental method in paving the way for studies on quantum effects in coupled metamaterials.

*corresponding author: liuhui@nju.edu.cn, zhusn@nju.edu.cn;

## I. Introduction

As a new type of artificial material, plasmonic metamaterials are not composed of natural atoms or molecules, but of micro-metallic/nano-metallic LC circuit resonators (also called meta-atoms or meta-molecules) [1–4]. When light interacts with metamaterials, oscillating currents will be induced in these LC resonators, leading to a strong magnetic plasmon resonance [1]. Therefore, high-frequency magnetism, which does not exist in natural materials, is obtained. It has been theoretically and experimentally confirmed that the dispersion of this medium strongly depends on the frequency of light, and that electric and magnetic fields are highly confined to subwavelength scales as the frequency of light obtains access to the resonance region of these LC resonators. Negative refraction and superlensing [5], cloaking [6, 7], and EIT (electromagnetically induced transparency) effect [4] all occur within such a frequency region, which leads to the worldwide attention paid to metamaterials.

In most of reported works up to now, the coupling interaction between meta-atoms are ignored and the effective properties of metamaterials can be seen as average effect of many single plasmonic resonators. However, in practical metmaterials, the coupling effect between meta-atoms should always exist, which is significant when the distance between resonators is small [8, 9]. After including coupling effect, the properties of metamaterials are quite different and complicated, which can be seen as a kind of "meta-solid". Similar to conventional solid-state matter, the concept of excitation has to be introduced to coupled metamaterials in accordance with the concepts of solid-state physics and quantum physics [10, 11].

Moreover，since the first demonstration of the plasmon-assisted entangled photons in perforated metal film has been reported, the quantum characteristics in plasmonic system and metamaterials have been continuously reported and have attracted more and more study interest for their potential applications in quantum information techniques [12–15]. All these require a profound understanding of the fundamental quantum properties of coupled metamaterials. Therefore, one can go forward to further study the interaction between coupled metamaterials and other materials.

In this paper, a therectical model is proposed to describe the quantum properties of coupled metamaterials in Section II. This model is used to investigate the interaction between coupled metamaterial and quantum dot in Section III. The damping of the compound system is also concerned in this part. After that, we present a realistic design of system composed by coupled metamaterial, a chain of nanosandwiched, and the PbS quantum dot in Section IV. The interaction between this system and PbS quantum dot are studies. And the stimulated emission radiation is calculated in this section. At last, the conclusion is given out in Section V.

**II. Quantum description of coupled metamaterials**

To study the quantum properties of coupled metamaterials, Hamiltonian combined with second quantization will be firstly established based our former classic Lagrangian model. Since the resonator can be seen as an L-C circuit, the coupled metamaterial was considered as a set of coupled resonators [2]. Without losing the universality, we investigated the one-dimensional case and considered only the nearest-neighbor coupling. The theoretical model of the coupled metamaterial is shown in Fig. 1. The Lagrangian of a coupled metamaterial is expressed as follows [8]:

$$\mathcal{L} = \sum_m \left[ \frac{L}{2}\dot{Q}_m^2 - \frac{1}{2C}Q_m^2 + \frac{M_h}{2}(\dot{Q}_m\dot{Q}_{m+1} + \dot{Q}_m\dot{Q}_{m-1}) - \frac{M_e}{2}(Q_mQ_{m+1} + Q_mQ_{m-1}) \right] \quad (1)$$

In this equation, $m$ is the number of unit cells; $L$ and $C$ are the inductor and the capacity of the single resonator, respectively; and $M_h$ and $M_e$ describe the magnetic and electric coupling between the unit cells, respectively. Meanwhile, $Q_m$ and $\dot{Q}_m$ correspond to the charge and current on the m-th unit cell, respectively. The terms formed by $\dot{Q}$ are the current-induced kinetic energies, and those consisting of $Q$ denote the charge-induced potential energy.

By using a Legendre transformation, $\mathcal{H} = \sum_m P_m\dot{Q}_m - \mathcal{L}$, we introduced the Hamiltonian of the coupled metamaterial, which plays a much more important role than the Lagrangian in solid-state physics. With $P_m = \partial\mathcal{L}/\partial\dot{Q}_m$ being the generalized

momentum that is correlated with $Q_m$, the Hamiltonian can be expressed as follows:

$$\mathcal{H}(Q_m, \dot{Q}_m) = \sum_m \left[ \frac{L}{2}\dot{Q}_m^2 + \frac{1}{2C}Q_m^2 + \frac{M_h}{2}(\dot{Q}_m\dot{Q}_{m+1} + \dot{Q}_m\dot{Q}_{m-1}) + \frac{M_e}{2}(Q_mQ_{m+1} + Q_mQ_{m-1}) \right]. \quad (2)$$

We carried out a Fourier transformation to this Hamiltonian and obtained the following:

$$\mathcal{H}(Q_k, \dot{Q}_{-k}) = \sum_k \left[ \frac{L}{2}\dot{Q}_k\dot{Q}_{-k} + M_h\cos(kd)\dot{Q}_k\dot{Q}_{-k} + \frac{1}{2C}Q_kQ_{-k} + M_e\cos(kd)Q_kQ_{-k} \right]. \quad (3)$$

In this equation, $d$ refers to the period of the metamaterial. We also used the Fourier expansion $Q_m = \frac{1}{\sqrt{M}}\sum_k Q_k e^{ikR_m}$. Since charge $Q_k$ has a canonically conjugate variable $P_k = \partial \mathcal{L}/\partial \dot{Q}_k = \left(\frac{L}{2} + M_h\cos(kd)\right)\dot{Q}_{-k}$, by using the Hamiltonian canonical equations $\dot{Q}_k = \partial \mathcal{H}/\partial P_k$ and $\dot{P}_k = -\partial \mathcal{H}/\partial Q_k$, we finally obtained the dispersion relation of the new excitation in the coupled metamaterial as follows:

$$\omega(k) = \omega_0 \sqrt{\frac{1 + \kappa_e \cos(kd)}{1 + \kappa_h \cos(kd)}}, \quad (4)$$

where $\omega_0^2 = 2\pi/LC$ is the resonant frequency of the unit cell itself and $\kappa_h = 2M_h/L$ and $\kappa_e = 2M_e C$ are the magnetic and electric coupling coefficient, respectively [8].

Considering the quantum condition, $\hat{Q}_m$ and $\hat{P}_m$ possess the commutation relation $[\hat{Q}_m, \hat{P}_m] = i\hbar$ [16, 17]. After some derivation, the commutator between $\hat{Q}_k$ and $\hat{P}_k$ can also be derived as $[\hat{Q}_k, \hat{P}_k] = i\hbar$. In the equation, we used the unitary condition $\frac{1}{M}\sum_m e^{i(k+k')md} = \delta_{k,-k}$. Bogoliubov transformation was performed to the Hamiltonian in Eq. (3) by introducing a set of creation and annihilation operators, $\hat{a}_k = U_k\hat{Q}_k + iV_k\hat{P}_{-k}$ and $\hat{a}_k^+ = U_k\hat{Q}_{-k} - iV_k\hat{P}_k$, with parameters $U_k = (\hbar)^{-1/2}\sqrt{\xi}$ and $V_k = (\hbar)^{-1/2}/\sqrt{\xi}$, and $\xi = \sqrt{[\frac{1}{2C} + M_e\cos(kd)][\frac{L}{2} + M_h\cos(kd)]}$ [16, 17]. The Hamiltonian of a coupled metamaterial in number representation can thus be obtained as follows:

$$\hat{\mathcal{H}} = \sum_k \left( \hat{a}_k^+ \hat{a}_k + \frac{1}{2} \right) \hbar \omega_k .\qquad(5)$$

In the equation, we obtained the quantum description of the excitation in a coupled metamaterial. The model of "quasi-particle" can be used here to give an institutive picture of quantum property of "meta-solid". $\hat{a}_k^+$ and $\hat{a}_k$ are the creating and annihilating operators, indicating the creation and destruction of "quasi-particle", respectively, with momentum $\hbar k$ in the metamaterial, and $\hat{a}_k^+ \hat{a}_k$ denotes number operator. The last term, $1/2 \hbar \omega_k$, is the energy of the vacuum state. The "quasi-particle" describes the electromagnetic resonance behavior of the entire solid-state-like metamaterial when there is coupling between unit cells. Once the coupling shrinks to zero, the metamaterial will return to the free-gas case, in which the model will simply correspond to the excitation of the unit cell itself. A possible experimental evidence of the quantum characteristic of metamaterials can be experimentally proved by measuring the second-order quantum coherence function $g^{(2)}(0)$ by using an attenuated-reflection set-up. For a quantum state, $|n\rangle$, $g^{(2)}(0)=1-1/n<1$ indicates a quantum property which can be measured directly in a practical experiment (see the third and fourth references in Ref. 15).

### III. Interaction between coupled metamaterial and quantum dot material

The metamaterial is an open artificial structured system that could be combined with any other inserted nature materials. Such a nature material could be a crystal with lattice vibration described by phonon; it could be a semiconductor quantum dot, in which exciton describes the properties of the electron-hole pairs; or it could also be ferromagnet or anti-ferromagnet, where magnon is the quantized spin wave among many other forms. The high confinement of field in a metamaterial can be used to enhance the coupling between photon and other quantum states in solid.

In this study, we took the compound system composed of metamaterial and quantum dot as an example. The interaction Hamiltonian of the system can be expressed as $\mathscr{H}_{Int} = \sum_{r} \boldsymbol{E} \cdot \boldsymbol{d}$; where $\boldsymbol{E}$ denotes the electric field in metamaterial, $\boldsymbol{d}$ refers to the dipole moment of exciton in the quantum dot, and the summation corresponds to all quantum dots in the system [18]. After some derivation, the quantized interaction Hamiltonian can be obtained as follows [19]:

$$\hat{\mathscr{H}}_{Int} = \hbar \sum_{k}[\mathbf{G}_{k}(\hat{a}_{k}^{+}\hat{\sigma}_{k}^{-} + \hat{a}_{k}\hat{\sigma}_{k}^{+})]. \tag{6}$$

In this equation, the coupling constant $\mathbf{G}_k$ is equal to $\sqrt{\int (\rho_2(\boldsymbol{\alpha}) - \rho_1(\boldsymbol{\alpha}))(\varphi_k(\boldsymbol{\alpha}) \cdot \boldsymbol{d}_{1,2})^2 d\boldsymbol{\alpha}^3} / \hbar$, which is of crucial importance in describing the interaction between photon and exciton. $\varphi_k(\boldsymbol{\alpha})$ is the eigenstate of photon corresponding to the electric field distribution with the energy normalized to $\hbar\omega_k/2$, with $\boldsymbol{\alpha}$ being the position of the quantum dot in the unit cell. $\rho_2$ and $\rho_1$ are the population densities of the two levels. The transition operators $\hat{\sigma}_k^{+}$ and $\hat{\sigma}_k^{-}$ indicate the creating and annihilating, respectively, of the quantum dots belonging to the whole system with momentum $\hbar\boldsymbol{k}$ [19]. The rotating wave approximation was used to eliminate the energy non-conversing terms.

Nearly all the quantum properties can be derived based on such interaction Hamiltonian between photon and exciton. Since the metamaterial is widely considered as a damping system, a set of quantum Langevin equations was introduced to describe the quantum motion of the Metamaterial-QD system, which can be expressed as follows:

$$\begin{cases} \dot{\hat{a}}_k = -i\omega\hat{a}_k - i\mathbf{G}_k\hat{\sigma}_k^{-} - \kappa\hat{a}_k/2 + \hat{F}(t); \\ \dot{\hat{\sigma}}_k^{-} = -i\omega\hat{\sigma}_k^{-} + i\mathbf{G}_k\hat{a}_k\hat{\sigma}_k^{z} - \gamma_{\perp}\hat{\sigma}_k^{-} + \hat{\varGamma}^{-}(t); \\ \dot{\hat{\sigma}}_k^{z} = 2i\mathbf{G}_k(\hat{\sigma}_k^{-}\hat{a}_k^{+} - \hat{a}_k\hat{\sigma}_k^{+}) + \gamma_{\parallel}(\varDelta_0 - \hat{\sigma}_k^{z}) + \hat{\varGamma}_z(t), \end{cases} \tag{7}$$

This considered the interaction with the reservoir. In the equation, $\kappa$, $\gamma_{\perp}$, and $\gamma_{\parallel}$ correspond to the damping of photon, $\hat{\sigma}_k^{-}$ and $\hat{\sigma}_k^{z}$. $\hat{F}(t)$, $\hat{\varGamma}^{-}(t)$, and $\hat{\varGamma}_z(t)$ indicate the reservoir forces [20], while $\varDelta_0$ is the fractional population difference. During the derivation, a Markovian approximation was employed. The quantum characteristic of the compound system can be sufficiently described by these equations.

**IV. The interaction between coupled nanosandwich chain and PbS quantum dots**

In this section, we designed a coupled metamaterial comprising a chain of nanosandwiches [21]. The geometry of a single nanosandwich is presented in Fig. 2(a). The metal is silver with a Drude-type permittivity of $\omega_p = 1.37 \times 10^{16} s^{-1}$ and $\gamma_m = 12.24 \times 10^{12} s^{-1}$ [8]. The substrate is glass, which has a refractive index of *1.5*. The middle layer of the nanosandwich is filled with semiconductor PbS quantum dot material with an emission wavelength of approximately *1,550 nm* and an electric permittivity of approximately *23*. If the quantum dots are densely packed in a vacuum, the effective permittivity of the middle layer would be calculated as *6.6* using the Maxwell-Garnett method [12]. Two cases were studied: a one-dimensional coupled metamaterial embedded in vacuum (Case I) and one embedded in quantum dot material (Case II). As shown in Figs. 2(b) and 2(c), the magnetic and electric couplings strongly depend on the period *d* between nanosandwiches. Generally, a smaller distance between resonators will produce stronger coupling. The dispersion properties of photon in both cases are shown in Figs. 2(d) and 2(e). For the smaller length of $a_x$, the electric field in Case II shows better confinement than that in Case I, leading to a stronger electric coupling between the nanosandwiches, as well as a broader excitation band [19]. In addition, the theoretical results from the Hamiltonian equation (green dotted line) are consistent with the simulation results, confirming the correctness of our processing method [22]. As seen in Figs. 2(d) and 2(e), the system had the lowest energy at the Brillouin zone center (*kd = 0*) and the highest energy at the edge of the Brillouin zone (*kd = π*). Therefore, according to the Hamiltonian in Eq. (3), $M_h$ and $M_e$ should both be positive, as were the results that we obtained. The eigenstate of the Meton $\varphi_k(\alpha)$ is presented in Fig. 3. The high-magnitude areas of the electric field indicate strong field confinement, leading to strong interaction between photon and exciton. Considering the maximum population inversion, we have $\rho_2 - \rho_1 \approx \rho$; where $\rho = \rho_2 + \rho_1$ is proportional to $r^{-3}$. Here, we took a moderate choice on the radius of the quantum dot as $r \approx 2.5\ nm$. The dipole moment was chosen to be $|d| = 1.9 \times 10^{-17}\ esu$ [12]. Then the coupling constant was calculated [19].

From the interaction Hamiltonian and coupling constant $\mathbf{G}_k$, the coupled kinetic equation for the emission processes of photon can be obtained. Under very strong optical pumped or electric pumped conditions, the energy absorbed by the quantum dots is very large and saturated. The number of exciton is much larger than the number of photon. We could assume that the number of exciton does not change, $\dot{\hat{\sigma}}_k^z \approx 0$ and only consider the change of photon. By using the adiabatic approximation and semi-classical approximation, the kinetic equation of photon $\mathcal{N}_{photon}^k = \langle \hat{a}_k^+ \hat{a}_k \rangle$, can be derived from Eq. (7) [20] as follows:

$$\frac{d\mathcal{N}_{photon}^k}{dt} = (\frac{2|\mathbf{G}_k|^2}{\gamma_\parallel} - \kappa)\mathcal{N}_{photon}^k + \frac{2|\mathbf{G}_k|^2}{\gamma_\parallel} \qquad (10)$$

The first term in the bracket indicates the stimulated emission, and the last term on the right-hand side of the equation is the spontaneous emission, which can be ignored for its small value. Since the homogeneous broadening of the quantum dot spectrum is much narrower than photon bandwidth [23], it is considered as a continuous radiation field and a narrow quantum-dot spectral case. Therefore, the stimulated emission rate must be integrated in a narrow frequency range, and the density of the state must be considered [19]. Finally, a stimulated emission rate $\mathscr{B}$ was obtained as $2|\mathbf{G}_k|^2 Md/v_g$, which is consistent with the results of Fermi's golden rule. In this equation, $M$ is the total number of unit cells in the metamaterial and $v_g$ is the group velocity derived from the dispersion relation of photon obtained above. Meanwhile, the damping term $\kappa$ can be considered as $\kappa = \gamma_{Meton}$, where $\gamma_{Meton} = 1/\tau_{Meton}$ represents the decay rate of photon. A sketch of the emission process is presented in Fig. 4(a).

The stimulated emission coefficient and lifetime of photon in both systems with different spacing are plotted in Figs. 4(b) and 4(c). When the spacing increases, the coupling between the unit cells decreases as well as photon's lifetime. From Figs. 4(b) and 4(c), we can see that the stimulated emission coefficients of photon in both cases were reduced along with the increase in spacing. With the larger number of available quantum dots, the interaction between the exciton and photon in Case II became stronger than that in Case I, leading to a larger $\mathscr{B}$. Furthermore, the stimulated emission coefficient $\mathscr{B}$ is dependent on the number of unit cells according to its expression. Thus, increasing $M$ can further enlarge the stimulated emission. The gain of the system is defined as $\Gamma = \mathscr{B}/\kappa - 1$. When $M$ is large, $\Gamma > 0$ can be easily obtained, and the amplification of photon by stimulated emission radiation occurs.

## V. Conclusion

In summary, we have studied the fundamental physics of coupled metamaterials. A quantum treatment is developed that enables people to investigate the quantum characteristics of coupled metamaterials, such as the interaction with quantum dots. A set of Langevin equations is presented to describe the quantum motion in such a compound system. These results can be used to treat the quantum properties in many fields such as nanolasers, LEDs, and solar cells. The merits of the Metamaterial-combined system not only include the broad bandwidth and rapid operation speed, but also the flexible structure for different frequencies or bandwidths and the strong coupling between photon and other quantum systems because of the high confinement of field. This is quite promising for the development of highly efficient single-photon detection and single-photon source devices. Furthermore, apart from the exciton, the interaction can also occur in photon-phonon, photon-magnon, and photon-biomolecule systems, leading to the potential application of metamaterials in optic-acoustics transfer devices, magneto-optics devices, and biosensors, among others.

**Acknowledge**

This work is supported by these National Programs of China (Grant Nos. 10704036, 10874081, 10534020, and Grant No. 2006CB921804).

**Caption:**

Fig. 1: (a) Model of coupled metamaterial; (b) Equivalent circuit.

Fig. 2: The geometry of a single nanosandwich is shown in (a). The coupling coefficients depending on $d$, with $a_x=215nm$, $a_y=300m$, $t=75nm$ and $h=50nm$ in case I and $a_x=160nm$, $a_y=300m$, $t=75nm$, and $h=50nm$ in case II are shown in (b) and (c), respectively. The dispersion relations of system for spacing $\Delta=d-a_x=50nm$ are shown in (d) and (e) for both cases, in which the yellow dotted lines are the light lines.

Fig. 3: The normalized electric field distribution at wavelength equal to *1550nm*, on the *z=0* plane of the nanosandwich in case I and case II are shown in (a) and (b), respectively.

Fig. 4: A sketch of the emission process is plotted in (a). The stimulated emission coefficient and life time of photon in case I and case II with different spacing are respectively plotted in (b) and (c).

**Figures:**

Fig. 1

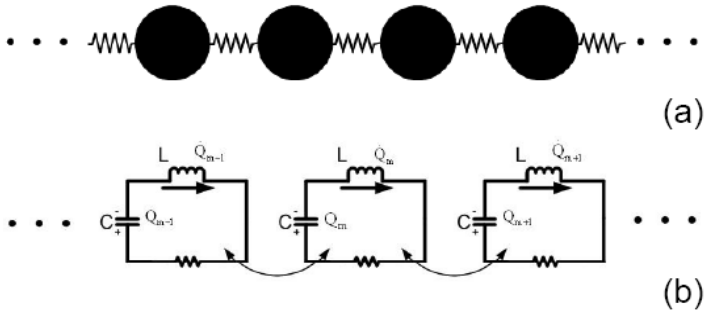

Fig. 2

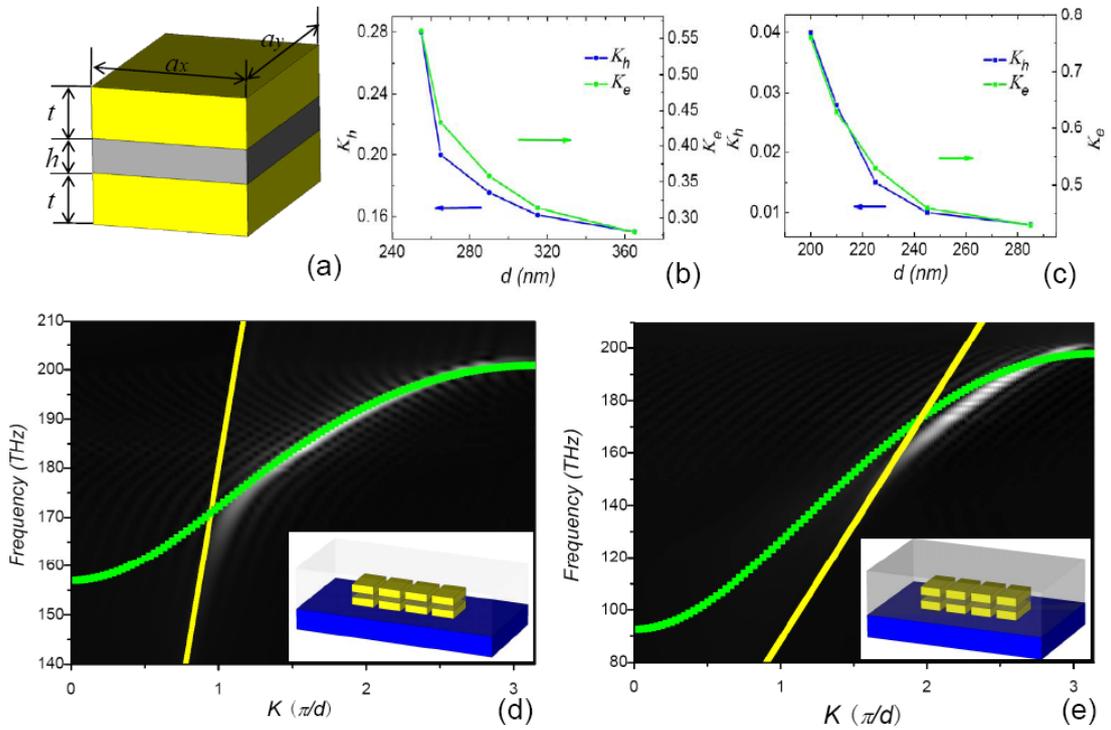

Fig. 3

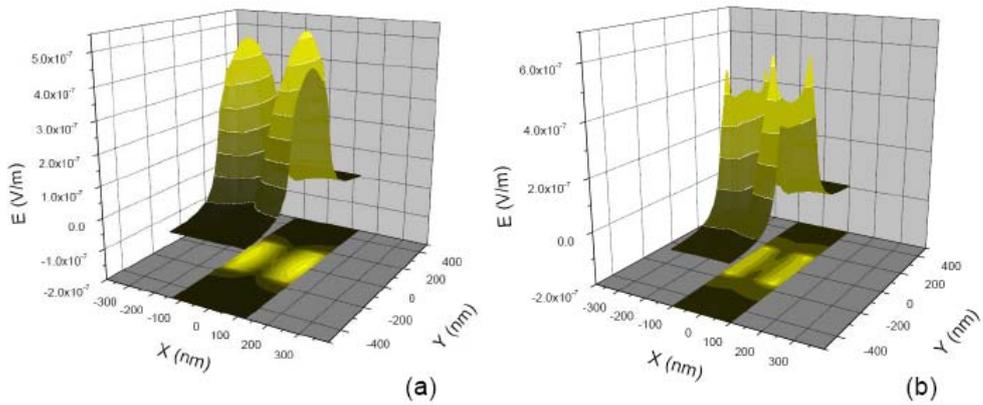

Fig. 4

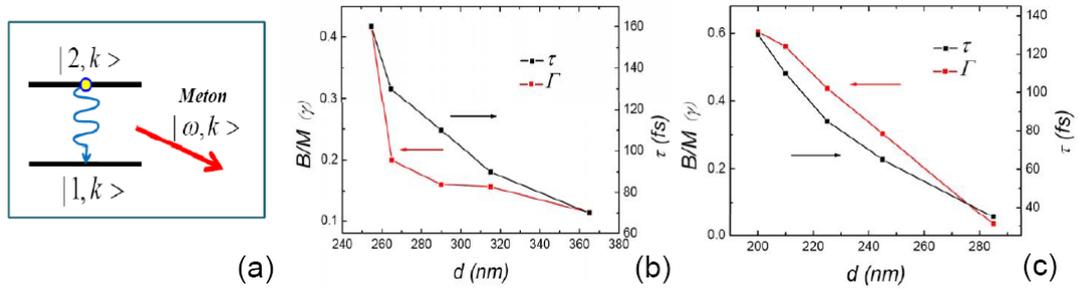

**Supplementary material: Part1**

**Fundamental of coupled metamaterial composed of metallic nanosandwiches**

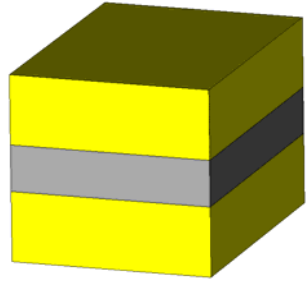

Fig. 1: The geometry of a single metallic nanosandwich.

A metallic nanosandwich is a good building block in metamaterial [1-3]. It can provide a magnetic resonance because of the opposite currents excited in the two slabs separated by a dielectric layer. The geometry of a single metallic nanosandwich is shown in Fig. 1.

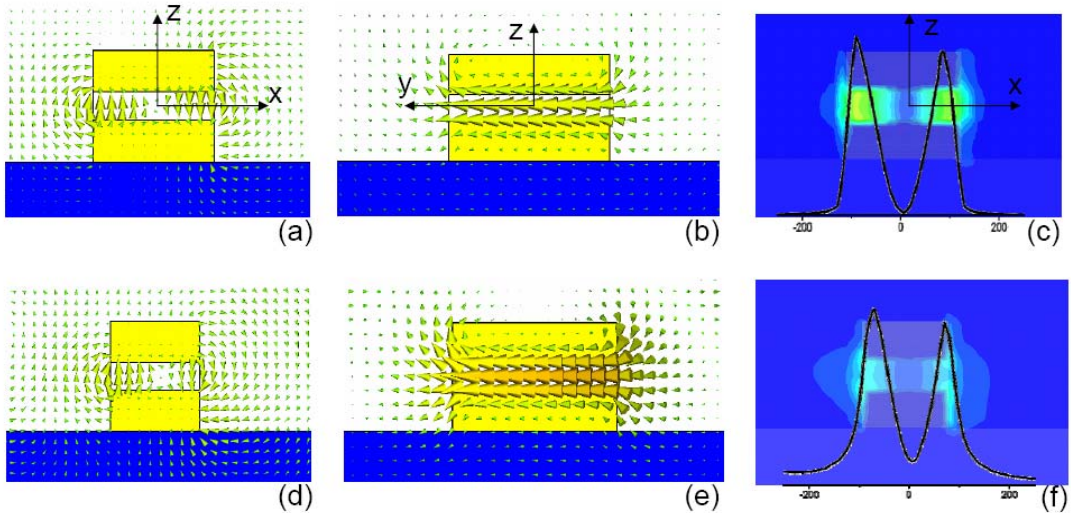

Fig. 2: (a)-(c) correspond to the electric field, magnetic field and electric energy distribution of a nanosandwich placed on the glass substrate embedded in vacuum at the magnetic resonant frequency. (d)-(f) belong to those nanosandwich embedded in PbS quantum dot material.

Fig. 2 shows the electric field and magnetic field of a nanosandwich placed on the glass substrate embedded in vacuum or PbS quantum dot material at the magnetic resonant frequency. It is evident that the magnetic resonance can be formed in such a metallic nanosandich. To show the field confinement of the nanosandwich, we plot the electric energy distribution of the nanosandwiches of both cases in Fig. 2(c) and (f), with the parameter of nanosandwiches as Fig. 2 (d) and (e) in the main text. From the Fig. 2(c), the electric energy rapidly reduces to zero outside the nanosandwich, while the speed of the case in Fig. 2(f) is much slower, which shows that the electric energy is confined better in (c) than that in (f) resulted from the larger length of $a_x$ and the refractive index contrast between the dielectric layer and outside surrounding in the vacuum case. It should be mentioned that for the strong coupling between unit cells in Case II, change of the band of Meton dependent on spacing is significant. Therefore, a tuning of $a_x$ from 160nm to 135nm, corresponding to $\Delta$ from 50nm to 150nm, is introduced to keep the Meton band covering 1550nm.

The dispersion relation of the Meton in metamaterial can be observed directly from the Fourier transformation map [4, 5]. Fig. 3 presents the dispersions of the one-dimensional metamaterials composed of a chain of metallic nanosandwiches with different spacings $\Delta = d - a_x$ between nanosandwiches.

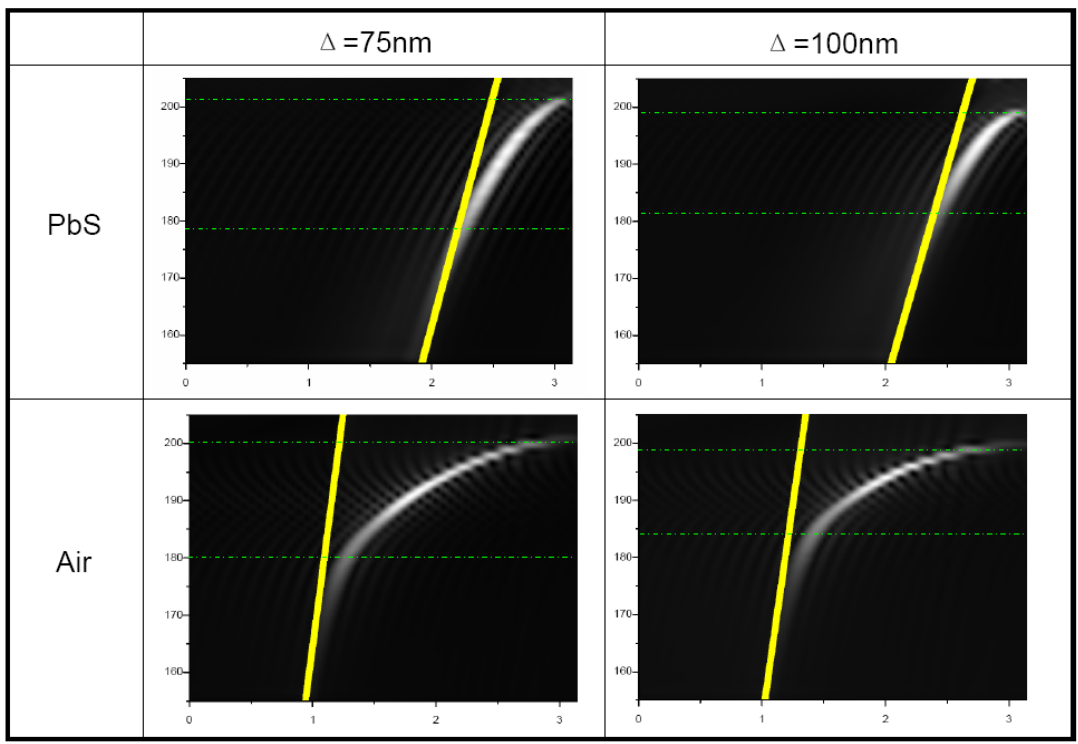

Fig. 3: The dispersion relations of one-dimensional metamaterial consist by a chain of metallic nanosandwiches with different spacings between in both cases. The yellow line is the light line. The green dash-dot lines remark the region of the band of Metons below the light line.

The dispersion curve of the Metons in metamaterial intersects with the light line. The part above the light line is leaky modes, which can easily exchange energy with the free modes in the outside surrounding. Therefore, they are quite weak that can hardly be seen in the Fourier transformation map. The part below the light line belongs to the localized modes of the one-dimensional metamaterial, which have the high field magnitude and can be easily observed from the Fourier map. For different spacing between nanosandwiches, the coupling effect will change. The coupling effect will reduce, when enlarging the spacing, leading to the shrinking of the bandwidth of the Metons in metamaterial. This is quite similar as the one-dimensional chain of atoms in the Solid State Physics [6]. The electric field and magnetic field distribution in the one-dimensional metamaterial are respectively shown in Fig. 4(a) and (b) by taking the case that metamaterial embedded in PbS quantum dot material with spacing being 75nm as an example. It is evident that the resonance belonging to the whole one-dimensional metamaterial is formed along the chain.

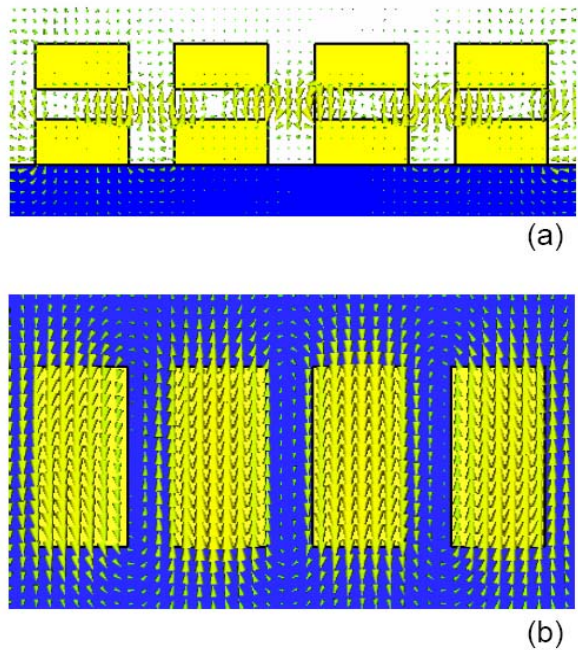

Fig. 4: The electric field and magnetic field distributions correspond to the metamaterial with spacing being 75nm embedded in PbS quantum dot material. The frequency is 193.5THz, namely 1550nm in wavelength.

**Supplementary material: Part2**

**Interaction between metamaterial and quantum dot material**

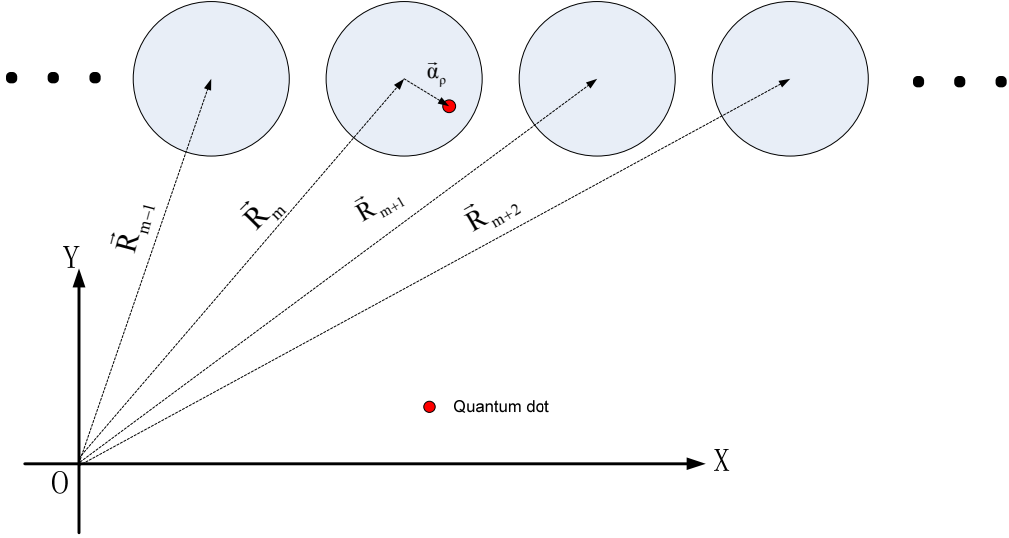

The interaction Hamiltonian of the compound system composed of metamaterial and quantum dot material has the following form

$$\mathcal{H}_{Int} = \sum_r -e\boldsymbol{d}(\boldsymbol{r}) \cdot \boldsymbol{E}(\boldsymbol{r}),  \quad (I)$$

where $e$ is the electron charge, $\boldsymbol{d}$ is electric dipole moment of a quantum dot [1]. The summation corresponds to all the quantum dots in the system.

We write the spatial electric field distribution of photon in the $m$-th unit cell as

$$\begin{aligned}\boldsymbol{E}(\boldsymbol{r}) &= \boldsymbol{E}(\boldsymbol{R}_m) = \frac{1}{\sqrt{V}} \int \boldsymbol{E}_{\boldsymbol{K}} e^{i\boldsymbol{K}\cdot\boldsymbol{r}} d\boldsymbol{K} = \frac{1}{\sqrt{M}} \sum_{k+G} \boldsymbol{E}_{k+G} e^{i(k+G)\cdot\boldsymbol{R}_m} \\ &= \frac{1}{\sqrt{M}} \sum_k \left( \sum_G \boldsymbol{E}_{k+G} e^{i(k+G)\alpha} \right) e^{ik\cdot\boldsymbol{R}_m} = \frac{1}{\sqrt{M}} \sum_k \boldsymbol{E}_k(\boldsymbol{\alpha}) e^{ik\cdot\boldsymbol{R}_m}\end{aligned} \quad (II)$$

where $\boldsymbol{R}_m$ and $\boldsymbol{\alpha}$ denote the center position of the $m$-th unit cell and the location of the quantum dot in this unit cell; $\boldsymbol{K}=\boldsymbol{k}+\boldsymbol{G}$, $\boldsymbol{K}$ is the wavevector of EM wave, $\boldsymbol{k}$ and $\boldsymbol{G}$ are the wave vector in the first Brillouin zone and the reciprocal lattice vector, and $M$ is the total number of unit cells. $\boldsymbol{E}_k(\boldsymbol{\alpha})$ corresponds to the electric field distribution of the unit cell with wave vector $\boldsymbol{k}$.

The quantized electric field can be expressed as

$$E(r) = E(R_m + \alpha) = \frac{1}{\sqrt{M}} \sum_k \varphi_k(\alpha)\left(\hat{a}_k e^{ikR_m} + \hat{a}_k^+ e^{-ikR_m}\right). \tag{III}$$

Here, $\varphi_k(\alpha)$ is the normalized electric field distribution with total energy of the unit cell normalized to $\hbar\omega_k/2$. The correction of this expression can be proved by substituting (III) into $\mathcal{H} = \int \varepsilon_0 d(\varepsilon\omega)/d\omega |E(r)|^2 + \mu_0 |H(r)|^2 \, dr/2$, where the magnetic field distribution $H(r)$ has the similar form as $E(r)$ with the normalized field distribution wrote as $\psi_k(\alpha)$. After some derivation, we can finally get the Hamiltonian as Eq. (3) in the paper.

The electric dipole moment can be written as $-ed(r) = -ed(R_m + \alpha) = \sum_{\varsigma,\xi} \hat{\sigma}_{\varsigma,\xi}^{m,\alpha} d_{\varsigma,\xi}$, in which $\varsigma$ and $\xi$ denote two levels of the exciton in the quantum dot [1]. Substituting (III) into (I), we get

$$\hat{\mathcal{H}}_{Int} = \frac{1}{\sqrt{M}} \sum_k \sum_m \sum_\alpha \sum_{\varsigma,\xi} \hat{\sigma}_{\varsigma,\xi}^{m,\alpha} d_{\varsigma,\xi} \cdot \varphi_k(\alpha)\left(\hat{a}_k e^{ikR_m} + \hat{a}_k^+ e^{-ikR_m}\right). \tag{IV}$$

After using a rotating-wave approximation to get rid of the energy non-conversing terms, we can obtain the interaction Hamiltonian as

$$\hat{\mathcal{H}}_{Int} = \frac{1}{\sqrt{M}} \sum_k \sum_m \sum_\alpha \varphi_k(\alpha) \cdot d_{1,2} \left(\hat{a}_k \hat{\sigma}_+^{m,\alpha} e^{ikR_m} + \hat{a}_k^+ \hat{\sigma}_-^{m,\alpha} e^{-ikR_m}\right). \tag{V}$$

The summation on $\alpha$ can be converted into an integral in the unit cell, by introducing a coupling constant

$$G_k = \sqrt{\int (\rho_2(\alpha) - \rho_1(\alpha))(\varphi_k(\alpha) \cdot d_{1,2})^2 d\alpha^3}/\hbar, \tag{VI}$$

where the integral is in the unit cell, $\rho_1(\alpha)$ and $\rho_2(\alpha)$ are the population of excitons on two levels. Considering the maximum population inversion, we have $\rho_2(\alpha) - \rho_1(\alpha) \approx \rho(\alpha)$, with $\rho(\alpha)$ being the concentration of quantum dot. A set of transition operators of quantum dots in the $m$-th unit cell,

$$\hat{\sigma}_+^m = \int \frac{\rho(\alpha) \varphi_k(\alpha) \cdot d_{1,2}^{m,\alpha}}{\sqrt{\int \rho(\alpha)(\varphi_k(\alpha) \cdot d_{1,2}^{m,\alpha})^2 d\alpha^3}} \hat{\sigma}_+^{m,\alpha} d\alpha^3, \tag{VII1}$$

$$\hat{\sigma}_-^m = \int \frac{\rho(\boldsymbol{\alpha})\varphi_k(\boldsymbol{\alpha}) \cdot \mathbf{d}_{1,2}^{m,\alpha}}{\sqrt{\int \rho(\boldsymbol{\alpha})\left(\varphi_k(\boldsymbol{\alpha}) \cdot \mathbf{d}_{1,2}^{m,\alpha}\right)^2 d\alpha^3}} \hat{\sigma}_-^{m,\alpha} d\alpha^3. \qquad (\text{VII2})$$

The transition indicating by these operators belong to all quantum dots in the unit cell. Then we can get

$$\hat{\mathcal{H}}_{Int} = \frac{1}{\sqrt{M}} \hbar \sum_k \sum_m \mathbf{G}_k \left( \hat{a}_k \hat{\sigma}_+^m e^{ikR_m} + \hat{a}_k^+ \hat{\sigma}_-^m e^{-ikR_m} \right). \qquad (\text{VIII})$$

Finally, we introduce a new set of transition operators of the whole system as

$$\hat{\sigma}_k^+ = \frac{1}{\sqrt{M}} \sum_m \hat{\sigma}_+^m e^{ikR_m}, \quad \text{and} \quad \hat{\sigma}_k^- = \frac{1}{\sqrt{M}} \sum_m \hat{\sigma}_-^m e^{-ikR_m}, \qquad (\text{IX})$$

corresponding to the transition belonging to the quantum dots of the entire quantum dot system. The simple interaction Hamiltonian can be obtained as

$$\hat{\mathcal{H}}_{Int} = \hbar \sum_k \mathbf{G}_k \left( \hat{a}_k \hat{\sigma}_k^+ + \hat{a}_k^+ \hat{\sigma}_k^- \right). \qquad (\text{X})$$

As for the stimulated emission, we have $\mathscr{B}_k = \frac{2|\mathbf{G}_k|^2}{\gamma_\parallel}$. For the continuous radiation field and narrow quantum dot spectral case, $\mathscr{B} = \int \mathscr{B}_k d\mathbf{k} \cdot \mathscr{D}$, in which $\mathscr{D} = \frac{Md}{\pi v_g}$ is the density of state, with $d$ being the spacing between unit cell, and $v_g$ corresponding to the group velocity of this one-dimensional system. Since the broadening of quantum dot spectrum have the form:

$$\tilde{g}(\nu,\nu_0) = \frac{\frac{\Delta \nu}{2\pi}}{(\nu - \nu_0)^2 + (\frac{\Delta \nu}{2})^2}, \qquad (\text{XI})$$

with $\Delta \nu = \gamma_\parallel / \pi$ and $\tilde{g}(\nu_0) = 2/\gamma_\parallel$ in $\mathscr{B}_k = |\mathbf{G}_k|^2 \tilde{g}(\nu_0) = \frac{2|\mathbf{G}_k|^2}{\gamma_\parallel}$.

After the integration, we have $\mathscr{B} = 2\pi |\mathbf{G}_k|^2 \mathscr{D}$, which is consistent with the Einstein relation. Inserting the density of state in one-dimensional chain, we have

$$\mathscr{B} = 2|\mathbf{G}_k|^2 \frac{Md}{v_g}. \qquad (\text{XII})$$

**Supplementary material: Part3**

It is an important issue of all the metallic nano/micro structures working in high frequency. A previous paper has considered the loss [1], in which, not the Lagrangian of metamaterial changes the form, but the Eular-Lagrange equation needs to add an additional term representing the loss. The general Lagrange equation is as follows.

$$\frac{d}{dt}\left(\frac{\partial L}{\partial \dot{Q}_i}\right) - \frac{\partial L}{\partial Q_i} = \Re_i \qquad \text{Eq.1}$$

Here, $Q_i$ and $\Re_i$ denote the generalized variable and the corresponding dissipation force, respectively. The loss term can write as $\Re_i = \gamma \dot{Q}_i$, where $\gamma$ is a loss coefficient [1, 2].

In this paper, we further investigate the Hamiltonian and Hamilton equations. The Hamilton equation will have a little change in

$$\dot{P}_k = -\frac{\partial H}{\partial Q_k} + \Re_k \qquad (2)$$

Here, a loss term $\Re_k$ is added, which is the Fourier transformation of $\Re_i$.

The loss effect finally results into a imaginary term of eigen frequency ω, whose reciprocal indicates the life time of meton. The life time of meton has been detailed investigated in the paper.

References:
7. H. Liu, et al., Phys. Rev. Lett. 97, 243902 (2006).
8. L.D. Landau and E.M. Lifshitz, Mechanics, 3$^{rd}$ ed, Pergamon Press, 1976.